# Kinetics and Dynamics of the S($^1$D$_2$) + H$_2$ → SH + H Reaction at Very Low Temperatures and Collision Energies

Coralie Berteloite,[1] Manuel Lara,[1*] Astrid Bergeat,[2,3] Sébastien Le Picard,[1] Fabrice Dayou,[1**] Kevin M. Hickson,[2,3] André Canosa,[1] Christian Naulin,[2,3] Jean-Michel Launay,[1†] Ian R. Sims[1]† and Michel Costes[2,3]†

[1] *Institut de Physique de Rennes, UMR 6251 du CNRS - Université de Rennes 1, Campus de Beaulieu, 35042 Rennes Cedex, France*

[2] *Université de Bordeaux, Institut des Sciences Moléculaires, 33405 Talence Cedex, France.*

[3] *CNRS UMR 5255, 33405 Talence Cedex, France.*

*\* Present adress: Departamento de Química Física Aplicada, Facultad de Ciencias, Universidad Autónoma de Madrid, 28049 Madrid, Spain*

*\*\* Present adress: LERMA, UMR 8112 du CNRS – Université Pierre et Marie Curie, Observatoire de Paris-Meudon, 92195 Meudon Cedex, France*

†To whom correspondence should be addressed. E-mail: jean-michel.launay@univ-rennes1.fr; ian.sims@univ-rennes1.fr; michel.costes@u-bordeaux1.fr.

We report combined studies on the prototypical S($^1$D$_2$) + H$_2$ insertion reaction. Kinetics and crossed-beam experiments are performed in experimental conditions approaching the cold energy regime, yielding absolute rate coefficients down to 5.8 K and relative integral cross sections to collision energies as low as 0.68 meV. They are supported by quantum calculations on a potential energy surface treating long range interactions accurately. All results are consistent and the excitation function behavior is explained in terms of the cumulative contribution of various partial waves.

Highly sophisticated physics experiments have recently made it possible to cool gas-phase molecules to very low temperatures and subsequently employ them as reagents of bimolecular chemical reactions [1]. One can cite the study of laser-cooled $Ca^+$ ions with velocity-selected $CH_3F$ molecules at collision energies equivalent to a few Kelvin [2] or the study of the KRb + KRb and K + KRb reactions performed at temperatures below 1 μK [3]. These ground-breaking experiments, realized under temperature and energy conditions where only a few partial waves contribute to reactivity, and where quantum effects can drastically change the reactive process, may furnish stringent tests for theory. However such reactions are not yet amenable to quantum mechanical calculations due to the large number of electrons to consider in the *ab initio* calculation of the potential energy surface (PES) and to the difficulties in performing an accurate quantum dynamical treatment. This letter reports the first experimental study of an elementary gas-phase reaction: $S(^1D_2) + H_2(X^1\Sigma_g^+, v = 0, j = 0,1) \rightarrow SH(X^2\Pi_i, v', j') + H(^2S_{1/2})$, with kinetics and crossed-beam experiments approaching the cold ($T < 1$ K) regime, supported by state-of-the-art quantum calculations [4].

Rate coefficients for the total removal of $S(^1D_2)$ by collisions with $H_2$ were measured using a CRESU apparatus [5,6] from $T = 298$ K down to $T = 5.8$ K, which is the lowest temperature ever achieved in kinetics experiments where laser cooling schemes are not applicable. Briefly, this experimental technique uses the supersonic expansion of a buffer gas through convergent-divergent Laval nozzles, each of them working under specific pressure conditions to cool the gas to a given temperature. A special double-walled nozzle was manufactured allowing pre-cooling to 77 K by liquid nitrogen, along with the reservoir upstream of the nozzle, to obtain a temperature of 5.8 K in the uniform supersonic flow. $S(^1D_2)$ atoms were produced in the supersonic flow by the laser photolysis of $CS_2$ at 193 nm. Variable concentrations of normal hydrogen, *n*-$H_2$, with an *ortho*:*para* ratio of 3:1, in large excess compared to $CS_2$ and hence $S(^1D_2)$, were added to the flow, ensuring that pseudo-first order kinetic conditions were maintained. At a given temperature and $n$-$H_2$ concentration, $S(^1D_2)$ atom decays were monitored by resonant laser-induced fluorescence (LIF) using the $3s^2 3p^4\ ^1D_2 \rightarrow 3s^2 3p^3 4s\ ^1D_2°$ transition at 166.67 nm. The $S(^1D_2)$ LIF decay traces (Fig. 1) were fitted by single exponential functions to extract the pseudo-first order rate coefficients $k_{1st}$ for the removal of $S(^1D_2)$ by $n$-$H_2$. The values of $k_{1st}$ were plotted as a function of the $n$-$H_2$ density to yield a straight line whose gradient corresponds to the second order rate coefficient $k(T)$ (Fig. 1). The measured rate coefficients are given in Table I plotted as a function of temperature in Fig. 2.

Relative integral cross sections (ICSs) were obtained from crossed-beam experiments [7]. $S(^1D_2)$ beams were generated by laser photolysis of $CS_2$ with 195 nm tuneable radiation at the throat of a cryocooled pulsed valve [8] operated with neon or helium carrier gas. The skimmed, pulsed $S(^1D_2)$ beam of velocity $v_S$ was crossed with a skimmed, pulsed $n$-$H_2$ beam of velocity $v_{H2}$ produced by a second cryocooled pulsed valve. The collision energy, $E_T$, or relative translational energy of the reagents with reduced mass $\mu$ and relative velocity $v_r$, was tuned by varying the beam intersection angle in the range $22.5° \leq \chi \leq 62.5°$: $E_T = \frac{1}{2} \mu v_r^2 = \frac{1}{2} \mu (v_S^2 + v_{H2}^2 - 2 v_S v_{H2} \cos\chi)$. The sulfur atom beams were characterized in the beam crossing region by (2+1) resonance-enhanced multiphoton ionisation (REMPI) using the $3s^2\ 3p^4\ ^1D_2 \rightarrow 3s^2\ 3p^3\ 4p\ ^1F_3$ transition at 288.18 nm. The precooled $H_2$ beams were probed when recording $(C^1\Pi_u, v = 2 \leftarrow X^1\Sigma_g^+, v = 0)$ (3+1) REMPI transitions around 289.5 nm [9] and were found to contain 75 % of molecules in $j = 1$ and 25 % in $j = 0$. The $H(^2S_{1/2})$ atoms produced by the reaction were detected by (1+1') REMPI via excitation of the Lyman-$\alpha$ $(^2S_{1/2} \rightarrow {}^2P°_J)$ transition at 121.567 nm followed by sequential ionisation with 364.7 nm photons. ICSs were deduced from $H^+$ intensities ($I_{H+}$) with the VUV laser tuned to the maximum of the H-atom Doppler spectrum while scanning the intersection angle with 2.5° increments. It is known that in such experiments the recorded signal may depend on i) the variation of the differential cross section with collision energy and ii) geometric as well as kinematic factors (variation of the collision volume with the intersection angle, laser irradiated volume and temporal width of the beams) related to the density-to-flux transformation [7]. All these factors were considered here. Doppler spectra of the recoiling H atom recorded at the two extremes of the collision energy range showed no marked difference. Therefore, effect i) was judged to be negligible and the ICS was simply taken as $(I_{H+}) / v_r$. Effect ii) was also unimportant in our operating conditions as shown by Figs 3 and 4 which display ICSs obtained in common collision energy ranges with different sets of beam velocities and thus with different beam intersection angles.

The $S(^1D_2) + H_2 \rightarrow SH + H$ system is a prototypical insertion reaction characterized by the presence of a deep well (4.13 eV) on its ground PES, the 1A' electronic state of $H_2S$, and by a small exoergicity ($\Delta H_0° = -0.292$ eV) [10]. Among the other four singlet PESs, 1A", 2A', 2A" and 3A' correlating with $S(^1D_2) + H_2$ reactants, only the 1A" PES connects adiabatically to the products. However, this 1A" PES which corresponds to the direct abstraction channel has a barrier of 0.43 eV in the collinear approach geometry [11]. It can thus be assumed that reaction occurs only on the 1A' PES at low energy. In a regime where the total energy is close to the reactant asymptote, a precise knowledge of the long-range interactions is crucial. The latter determines the amount of incoming flux reaching the short-range region where strong chemical forces act and rearrangement occurs [12]. Short-range interactions, described using a currently existing *ab initio* PES [13] were thus complemented with new accurate calculations of the long-range interactions. They include the interactions between the permanent quadrupole moments of the reactants (varying as $R^{-5}$) and the dispersion interactions between the dipole-induced moments of the two species (varying as $R^{-6}$). Interestingly, a significant anisotropic quadrupole-quadrupole contribution should lead to important reorientation effects in the entrance channel particularly at low energies. The hyperspherical quantum reactive scattering method [14] with slight modifications to allow for anisotropic effects has been applied to this PES to calculate the observables measured in the experiments. Rate coefficients corresponding to $n$-H$_2$ are presented in Fig. 2. Individual integral cross sections for H$_2$($j$ = 0) and H$_2$($j$ = 1) along with several partial wave contributions are displayed in Fig. 3.

As can be seen in Fig. 2, the experimental and theoretical rate coefficients remain very high down to the lowest temperatures as expected for an exothermic reaction without a barrier. Excellent agreement is obtained at room temperature between the experimental rate coefficient and the only previous measurement [15]. There is good agreement between the experimental and theoretical values of $k(T)$ regarding both their absolute values (note that no scaling has been employed) and the overall trend of a monotonic increase in $k(T)$ with $T$. It should be noted, though, that the kinetics experiments probe the total removal of S($^1$D$_2$) by $n$-H$_2$, $i.e.$ reaction plus relaxation S($^1$D$_2$) + H$_2$ → S($^3$P$_{0,1,2}$) + H$_2$, and thus yield an upper limit to the true reactive rate coefficient. Theory treats the collision adiabatically however. Intersystem crossing between the singlet and triplet PESs is ignored and the relaxation process is not considered. A previous theoretical trajectory surface-hopping study at higher energies indicated that the electronic quenching process may play a significant role in the total removal of S($^1$D$_2$) [16]. Nevertheless, the current theory accounts well for the total number of complexes formed in the collision, which finally decompose to give either products $via$ reaction or ground-state reactants $via$ electronic quenching. This explains the agreement of the theoretical rates with the experimental total removal rates, where both possibilities exist. Concerning the observed inversion between the experimental and theoretical rate coefficients at 5.8 K, the theoretical result at such a low temperature becomes highly sensitive to even small inaccuracies in the 1A' PES and also on the effect of non-adiabatic couplings between this PES and the 2A' and 3A' PESs which have been neglected in the present work.

The absolute theoretical ICSs for $H_2(j = 0$ and $1)$ and the relative experimental ICSs displayed in Figs. 3 and 4 exhibit the general form characteristic of an exoergic barrierless reaction, with an overall Langevin capture-type behavior, *i.e.* a cross section which increases with decreasing collision energy. The ICS theoretical curve matching the experimental conditions, *i.e.* 75 % of collisions with $H_2(j = 1)$ and 25 % of collisions with $H_2(j = 0)$ and convoluted over the collisional energy spread is also shown in Fig. 3c and 4 for comparison with the crossed-beam results. The lower experimental limit $E_T = 0.68$ meV is the lowest collision energy ever attained in crossed-beam experiments between neutral species. The upper experimental limit $E_T = 30$ meV reaches the collision energy range covered by previous crossed-beam experiments [17]. Superimposed on the general Langevin behavior, subtle undulations originating from the highly structured reaction probability [18] are observed in the theoretical excitation functions. This is more pronounced for $H_2(j = 0)$ (Fig. 3a) than for $H_2(j = 1)$ (Fig. 3b). All of these oscillations are associated with the opening of new partial waves, but some of them with high total angular momentum $J$ occur at collision energies below the height of the centrifugal barrier and therefore correspond to shape resonances [19]. The weighting of ICSs by the 3:1 *ortho*:*para* ratio scrambles the individual patterns. However, the experimental excitation function does reveal undulations in the low energy region which culminate in a pronounced plateau between 1 and 2 meV.

The agreement between theoretical and experimental ICSs is good. A rate coefficient calculated on the basis of the experimental ICSs will exhibit the same temperature dependence as the theoretical one (see Fig. 2). The agreement between the crossed-beam and the CRESU results is thus good between 298 and 23 K. A direct comparison at the lowest temperature is not possible since 56 % of collisions of the thermal distribution of velocities at 5.8 K occur at energies below 0.68 meV and would require extrapolation of the ICSs to lower energies.

This work has set new low temperature and collision energy limits for combined experimental and theoretical studies of gas-phase reactive collisions for which laser cooling schemes are not applicable. The use of a para-$H_2$ beam in the crossed-beam experiments is crucial to reveal the partial wave structure and the possible occurrence of resonances and such experiments on $S(^1D_2) + H_2(j = 0)$ are now in progress. Other systems will be investigated, in particular the benchmark $F + H_2$ reaction for which considerable experimental and theoretical interest exists.

**Acknowledgments** This research is supported by the Agence Nationale de la Recherche under contract ANR-BLAN-2006-0247 Cold Reactions between Neutral Species which provided fellowships for CB, ML and FD, and by the Conseil Régional d'Aquitaine under contract 2007.1221 Collisions Réactives Froides. The authors acknowledge Ian W. M. Smith for his comments on the manuscript. The Bordeaux group wish to express their thanks to Uzi Even for all his advice and to Nachum Lavie for the reduced delays in the construction of the low temperature pulsed valves.

**TABLE I.** Rate coefficients $k$ for the removal of S($^1$D$_2$) by $n$-H$_2$ measured by the CRESU technique.

| $T$ / K | [He] / $10^{16}$ cm$^{-3}$ | [$n$-H$_2$] / $10^{13}$ cm$^{-3}$ | $k$ / $10^{-10}$ cm$^3$ s$^{-1}$ |
|---|---|---|---|
| 5.8 | 14.2 | 3.4 – 32.2 | 0.72 ± 0.13[a] |
| 5.8 | 14.2 | 4.08 – 40.2 | 0.565 ± 0.07 |
| 23 | 4.73 | 8.8 – 67.7 | 1.52 ± 0.16 |
| 49 | 10.4 | 5.3 – 52.5 | 1.50 ± 0.15 |
| 123 | 12.7 | 0 – 35.7 | 1.94 ± 0.20 |
| 298 | 29.2 | 1.3 – 132 | 2.07 ± 0.21 |

[a] Uncertainties include systematic (estimated at *ca*. 10%) and statistical errors (evaluated from the second order plots at 95% confidence).

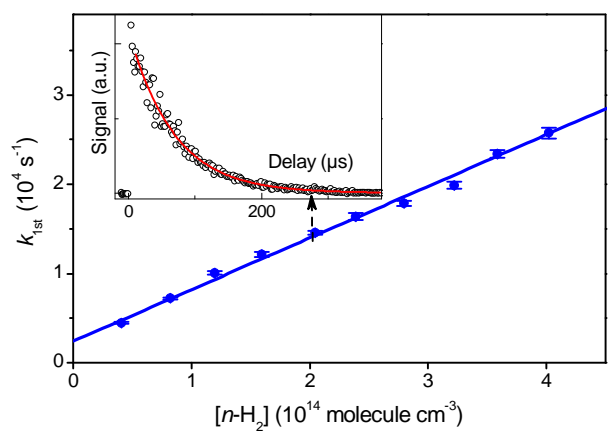

**FIG. 1.** Pseudo-first-order rate coefficients plotted against the concentration of $n$-$H_2$ at 5.8 K. The inset shows the fitted decays of the S($^1D_2$) laser-induced fluorescence observed in mixtures containing $2.04 \times 10^{14}$ molecule cm$^{-3}$ $n$-$H_2$.

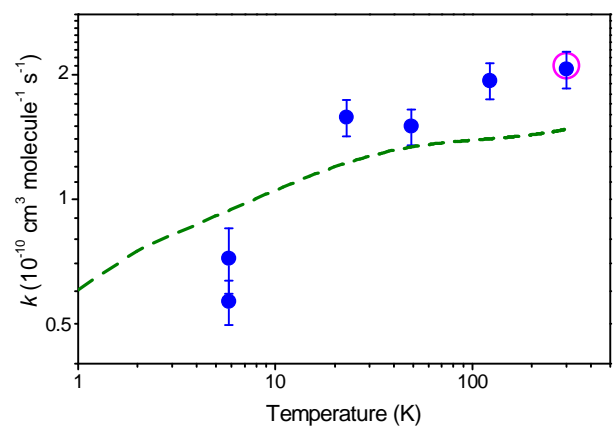

**FIG. 2.** A comparison of experimental (filled circles) and theoretical (dashed line) rate coefficients for $S(^1D_2)$ + $n$-$H_2$ collisions. The only previous experimental result at 300 K (open circle) is also shown [15].

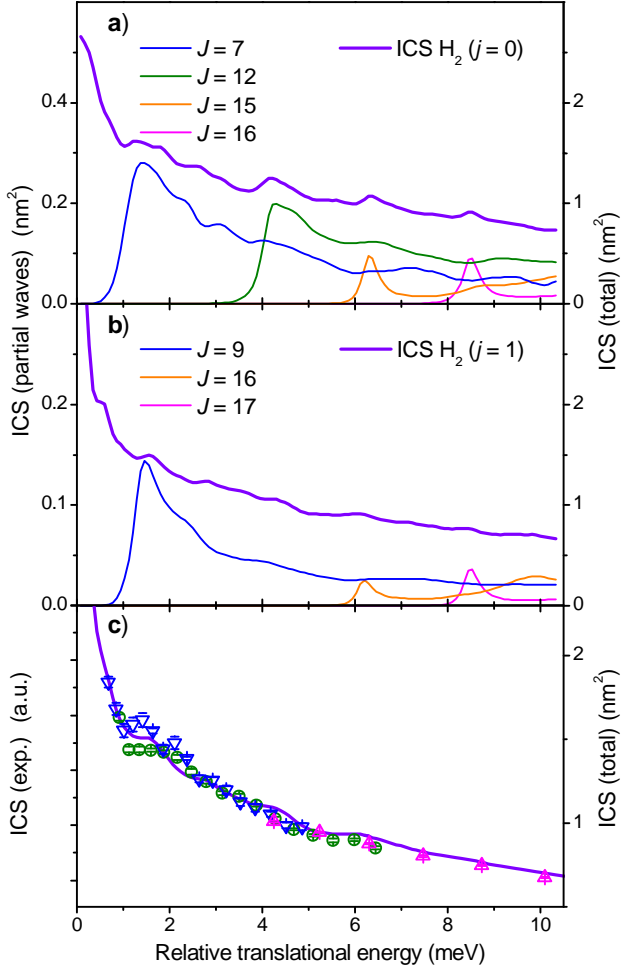

**FIG. 3.** Integral cross sections (ICS) of the $S(^1D_2) + H_2$ reaction.

Theoretical ICSs computed for (a) $H_2$ ($j = 0$) and (b) $H_2$ ($j = 1$): total ICS (right vertical scale); some of the partial waves (left vertical scale) are also displayed for given values of total angular momentum $J$. The experimental measurements (c), obtained for 3 different sets of beam velocities, which are normalized in the overlapping energy range, are displayed in arbitrary units (left vertical scale) as symbols under the following beam conditions: open triangles (▽): $v_S = v_{H2} = 678$ m s$^{-1}$; open circles (o): $v_S = v_{H2} = 780$ m s$^{-1}$; open triangles (△): $v_S = v_{H2} = 1685$ m s$^{-1}$. The error bars correspond to 95% confidence limits. The high energy part between $E_T = 10$ meV and $E_T = 30$ meV is not represented for clarity (see Fig. 4). The solid curve refers to the theoretical ICSs calculated as a combination of 3:1 weighted contributions due to *ortho*:*para* absolute integral cross sections displayed above, convoluted with a Gaussian distribution of collision energies to account for velocity and crossing angle spread in the experiments ($\Delta E_T/E_T \approx 7\%$ half-width at 1/e).

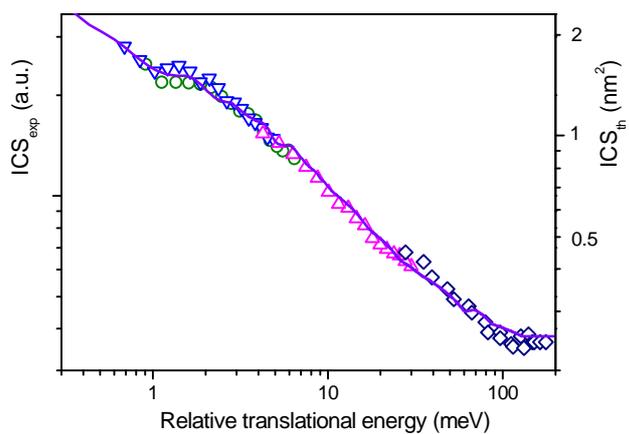

**FIG. 4.** Integral cross sections (ICS) of the $S(^1D_2)$ + $H_2$ reaction displayed on the whole energy range.

Same symbols as in Fig. 3; diamonds (◇) previous data [17].